\documentclass{pasj00}
\draft

\begin{document}
\SetRunningHead{Kohno et al.}{ASTE Observations of Dense-Gas Bar in NGC 986}
\Received{}
\Accepted{}

\title{ASTE CO(3--2) Observations of the Southern Barred Spiral Galaxy NGC 986:
a Large Gaseous Bar Filled with Dense Molecular Medium}

%
 \author{%
   Kotaro \textsc{Kohno},\altaffilmark{1}
   Tomoka \textsc{Tosaki},\altaffilmark{2}
   Rie \textsc{Miura},\altaffilmark{3,4}
   Kazuyuki \textsc{Muraoka},\altaffilmark{1} 
   Tsuyoshi \textsc{Sawada},\altaffilmark{2} \\
   Kouichiro \textsc{Nakanishi},\altaffilmark{2}
   Nario \textsc{Kuno},\altaffilmark{2} 
   Takeshi \textsc{Sakai},\altaffilmark{2}
   Kazuo \textsc{Sorai},\altaffilmark{5}
   Kazuhisa \textsc{Kamegai},\altaffilmark{1} \\
   Kunihiko \textsc{Tanaka},\altaffilmark{1} 
   Takeshi \textsc{Okuda},\altaffilmark{1}\altaffilmark{$*$}
   Akira \textsc{Endo},\altaffilmark{1,3}
   Bunyo \textsc{Hatsukade},\altaffilmark{1} \\
   Masahiro \textsc{Sameshima},\altaffilmark{1} 
   Hajime \textsc{Ezawa},\altaffilmark{3}
   Seiichi \textsc{Sakamoto},\altaffilmark{3}\altaffilmark{$\dagger$}
   Takeshi \textsc{Kamazaki},\altaffilmark{3} \\
   Nobuyuki \textsc{Yamaguchi},\altaffilmark{3}
   Juan \textsc{Cortes},\altaffilmark{3,6}
   Yoichi \textsc{Tamura},\altaffilmark{3,4}
   Masayuki \textsc{Fukuhara},\altaffilmark{3,4} \\
   Daisuke \textsc{Iono},\altaffilmark{3}\altaffilmark{$\ddagger$} and
   Ryohei \textsc{Kawabe},\altaffilmark{3}
}

 \altaffiltext{1}{Institute of Astronomy, The University of Tokyo, 2-21-1 Osawa, Mitaka, Tokyo 181-0015}
 \email{(KK) kkohno@ioa.s.u-tokyo.ac.jp}
 \altaffiltext{2}{Nobeyama Radio Observatory, Minamimaki, Minamisaku, Nagano 384-1805}
 \altaffiltext{3}{National Astronomical Observatory of Japan, 2-21-1 Osawa, Mitaka, Tokyo 181-8588}
 \altaffiltext{4}{Department of Astronomy, The University of Tokyo, Hongo, Bunkyo-ku, Tokyo 113-0033}
 \altaffiltext{5}{Division of Physics, Graduate School of Science, Hokkaido University, Sapporo 060-0810}
 \altaffiltext{6}{Departamento de Astronomia, Universidad de Chile, Casilla 36-D, Santiago, Chile}

 \altaffiltext{$*$}{Present address is National Astronomical Observatory of Japan, \\
 2-21-1 Osawa, Mitaka, Tokyo 181-8588} 
 \altaffiltext{$\dagger$}{Present address is Institute of Space and Astronautical Science, \\
 Japan Aerospace Exploration Agency, 3-1-1 Yoshinodai, Sagamihara, Kanagawa 229-8510}
 \altaffiltext{$\ddagger$}{Present address is Institute of Astronomy, The University of Tokyo, \\
 2-21-1 Osawa, Mitaka, Tokyo 181-0015}


\maketitle

\begin{abstract}
We present CO(3--2) emission observations 
toward the $3' \times 3'$ (or 20 $\times$ 20 kpc at a distance of 23 Mpc) region of
the southern barred spiral galaxy NGC 986
using the Atacama Submillimeter Telescope Experiment (ASTE). 
This effort is a part of our on-going 
extragalactic CO(3--2) imaging project ADIoS 
(ASTE Dense gas Imaging of Spiral galaxies).
Our CO(3--2) image revealed the presence of 
a large (the major axis is 14 kpc in total length) gaseous bar
filled with dense molecular medium
along the dark lanes observed in optical images.
This is the largest ``dense-gas rich bar'' known to date.
The dense gas bar discovered in NGC 986 could be a huge reservoir
of possible ``fuel'' for future starbursts in the central region, 
and we suggest that the star formation in the central region of NGC 986 
could still be in a growing phase.
We found a good spatial coincidence between the overall distributions
of dense molecular gas traced by CO(3--2) 
and the massive star formation depicted by H$\alpha$.
The global CO(3--2) luminosity $L'_{\rm CO(3-2)}$ of NGC 986
was determined to be $(5.4 \pm 1.1) \times 10^8$ K km s$^{-1}$ pc$^2$. 
The CO(3--2)/CO(1--0) integrated intensity ratio was found to be 0.60 $\pm$ 0.13
at a spatial resolution of $44''$ or 5 kpc, and a CO(3--2)/CO(2--1) ratio was
0.67 $\pm$ 0.14 at a beam size of $\sim$25$''$ or $\sim$2.8 kpc.
These line ratios suggest moderate excitation conditions of CO lines 
($n_{\rm H_2} \sim 10^{3-4}$ cm$^{-3}$)
in the central a few kpc region of NGC 986. 
\end{abstract}

\section{Introduction}

The dense molecular medium is one of the indispensable components 
to understand the star formation law in galaxies. 
This is because stars are formed from dense molecular cores,
not from the diffuse envelopes of giant molecular clouds (GMCs).
In fact, extragalactic observations of millimeter-wave HCN(1--0) emission,
which is a tracer of high density molecular medium
($n_{\rm H_2} \sim 10^5$ cm$^{-3}$)
due to its high permanent dipole moment ($\mu$ = 3.0 Debye),
demonstrate that the correlation between HCN(1--0) 
and far-infrared (FIR) luminosities is 
better than that between CO(1--0) and FIR luminosities
(\cite{sdr1992}; \cite{gao2004a}).
However, the weakness of HCN emission 
(typically 1/10 in $T_{\rm b}$ of CO 
in the central regions of galaxies, and 1/30 - 1/50 of CO
in the disk regions of the Galaxy (\cite{helfer1997b})
and galaxies (\cite{helfer1993}; \cite{kohno1996}; \cite{kohno1999}; \cite{kohno2003};
\cite{gao2004b}) often prevents us from obtaining
large-scale maps of the dense molecular medium through HCN observations.

In order to understand the global distribution of dense molecular medium in galaxies,
we have conducted an extragalactic CO(3--2) imaging survey of nearby
spiral galaxies, ADIoS (ASTE Dense gas Imaging of Spiral galaxies).
The current sample galaxies of the ADIoS project are listed in Table 1.
The submillimeter-wave CO(3--2) emission 
is another tracer of dense gas because 
its Einstein A coefficient is proportional to $\nu^3$;
therefore, the critical density of CO(3--2) emission is higher than that of CO(1--0) 
by a factor of $\sim3^3$, i.e., $n_{\rm H_2} \sim 10^4$ cm$^{-3}$. 

The Atacama Submillimeter Telescope Experiment (ASTE; \cite{ezawa2004}), 
a new project to operate a 10 m telescope in the Atacama Desert of northern Chile, 
provides us with an ideal opportunity 
to generate large-scale maps of CO(3--2) emission 
because of the high beam efficiency of the telescope,
a low system noise temperature 
due to the good receiver system and atmospheric conditions at the site, 
and its efficient on-the-fly (OTF) mapping capability (\cite{sawada2008}).

Here, we present the CO(3--2) images of the southern CO luminous spiral galaxy NGC 986
as the initial result of the ADIoS project by using OTF mapping.
NGC 986 is a nearby ($D$ = 23.2 Mpc; \cite{tully1988}) barred spiral galaxy 
accompanied with an outer pseudo ring (i.e., (R$'_1$)SB(rs)b, \cite{buta1995}).
It is a member of the IRAS bright galaxy sample (\cite{sanders2003}); however,
its FIR luminosity is rather moderate ($L_{\rm FIR} = 3.5 \times 10^{10} L_\odot$)
\footnote{This $L_{\rm FIR}$ is rescaled to the value for the distance we assumed here.}.
The nucleus is classified as HII (\cite{vv1986}), and
extended massive star forming regions along the bar have been imaged in
H$\alpha$ (\cite{hameed1999}; \cite{koopmann2006})
and mid-infrared continuum (\cite{dale2000}; see also \cite{forster2004}).
The central region of NGC 986 is very rich in interstellar medium (ISM), and
there are many reports on the observations of 
various atomic lines in the near-infrared (NIR) to FIR regions 
(\cite{kawara1987}; \cite{kawara1989}; \cite{thornley2000}; \cite{malhotra2001})
and molecular lines at millimeter wavelengths 
(\cite{aalto1991}; \cite{aalto1995}; \cite{elfhag1996}).
In fact, NGC 986 is one of the brightest galaxies in terms of CO(1--0) emission 
in the sample of \citet{elfhag1996}, 
thereby rendering this galaxy as an ideal target for our ADIoS project.
It should be noted that all the previous CO observations of NGC 986 were conducted 
just toward the central position and no CO map of NGC 986 has been published yet.

\section{Observations and Data Reduction}

The CO(3--2) observations towards NGC 986 were conducted using the ASTE
on September 14 and 15, 2006. The total time for observation was 10 hours.
The mapped region was $3' \times 3'$ ($20 \times 20$ kpc),
covering almost the entire region of the optical disk.
The half-power beam width (HPBW) of the ASTE 10 m dish is 22$''$
at this frequency; this corresponds to 2.5 kpc at a distance of 23 Mpc.

The front end was a cartridge-type cooled SIS mixer receiver for the DSB operation, 
SC345 (\cite{kohno2005}; \cite{muraoka2007}).
The back end was a digital autocorrelator system, MAC (\cite{sorai2000mac}),
which is comprised of four banks of a 512 MHz wide spectrometer with 1024 spectral channels each.
This arrangement provided a velocity coverage of 440 km s$^{-1}$ 
with a velocity resolution of 0.43 km s$^{-1}$.
The observations were made remotely from an ASTE operation room of NRO
using a network observation system N-COSMOS3 developed 
by the National Astronomical Observatory of Japan (NAOJ) (\cite{kam05}).

OTF mapping was performed along two different directions 
(i.e., scans along the RA and Decl.\ directions),
and these two data sets were co-added by the Basket-weave method
(\cite{emerson1988}) in order to remove effects of scanning noise.
At the end of each OTF scan, an off-source position 
($5'$ offset from the map center in the azimuth)
was observed to subtract sky emission.

We observed the CO(3--2) emission of O-Ceti (Mira) every 2 hours 
in order to monitor the stabilities
of the pointing accuracy and main beam efficiency ($\eta_{\rm mb}$) of the ASTE 10 m dish. 
The pointing accuracy was found to be better than $\sim 2''$ r.m.s.\,
and $\eta_{\rm mb}$ was estimated to be 0.6 during the observing runs.
The absolute error of the CO(3--2) amplitude scale was $\sim$ $\pm$ 20~\%, 
mainly due to the variation in the beam efficiency.

The data reduction was made using the software package {\it NOSTAR}, 
which comprises tools for OTF data analysis developed by NAOJ (\cite{sawada2008}). 
The raw data were regridded to 8$''$ per pixel, giving an effective spatial
resolution of approximately $25''$ (or 2.8 kpc).
Linear baselines were removed from the spectra. 
We binned the adjacent channels to a velocity resolution of 10 km s$^{-1}$ 
at the frequency of CO(3--2).
The resultant r.m.s.\ noise level (1 $\sigma$) was around 25 mK 
on the $T_{\rm mb}$ scale
or 1.5 Jy beam$^{-1}$ at a beam size of $25''$ (HPBW).
It is noteworthy that at this frequency and beam size, 
a brightness temperature of 1 K on the $T_{\rm mb}$ scale 
corresponds to a flux density of 60.7 Jy beam$^{-1}$ .

After producing a 3D data cube, we analyzed it using AIPS.
The 0th and 1st moment maps were computed 
using a clip level of 50 mK (or 2 $\sigma$) by the AIPS task MOMNT.
A three-channel Hanning smoothing along the velocity axis was
applied in order to mitigate the contribution from noise.

\section{Results}

\subsection{Channel maps, spectrum, and moment maps}

\begin{figure}
  \begin{center}
    \FigureFile(80mm,80mm){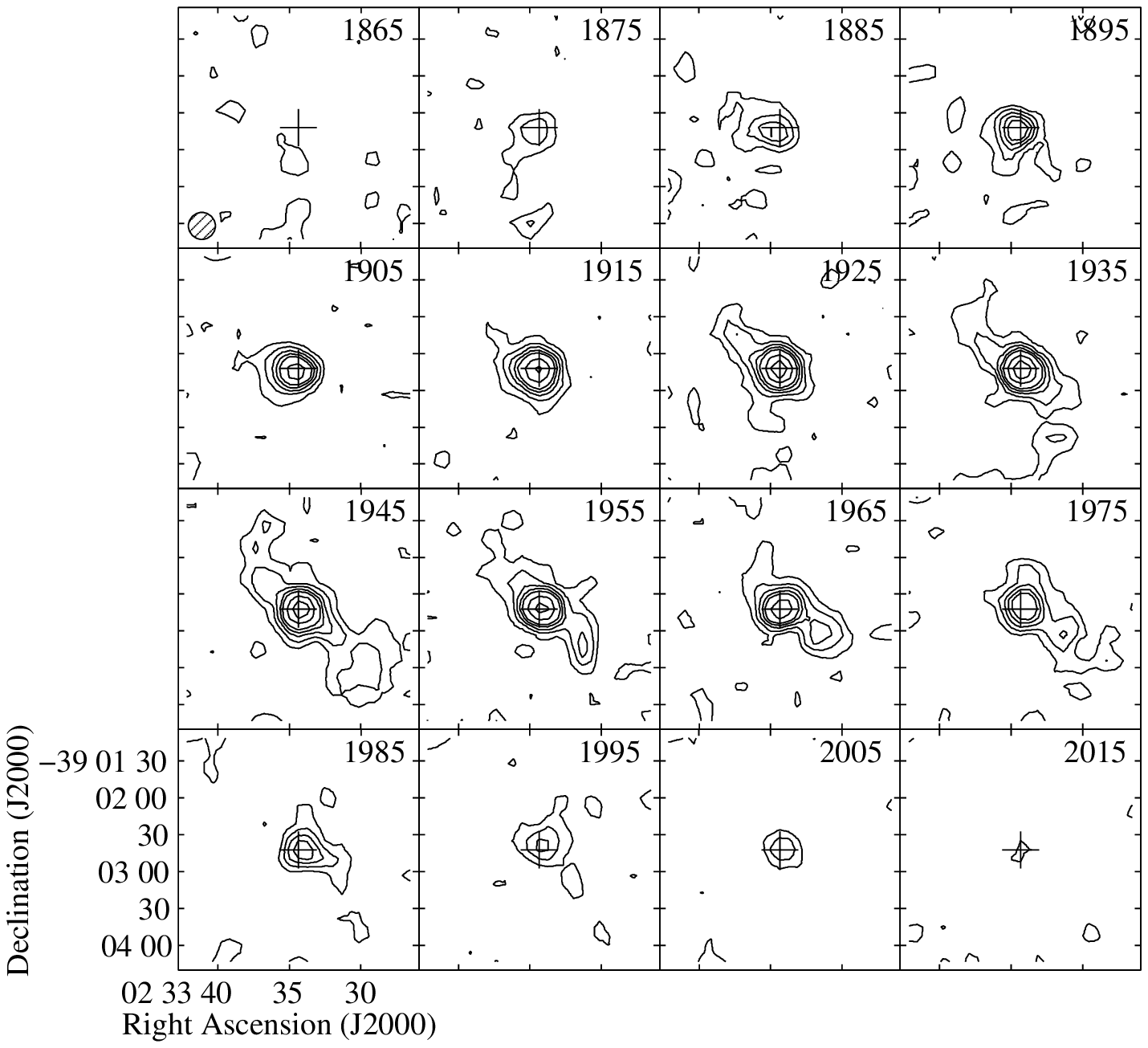}
  \end{center}
\caption{
Velocity channel maps of the CO(3--2) emission in NGC 986 obtained by ASTE.
Contour levels are 2, 4, 6, 8, 10, 15, 20, and 25 $\sigma$, where
1 $\sigma$ = 25 mK on the $T_{\rm mb}$ scale or 1.5 Jy beam$^{-1}$. 
Each map is labeled by the LSR velocity in km s$^{-1}$.
The central cross in each panel indicates the position of the nucleus
(defined by the peak position of the NIR continuum from 2MASS/NED).
}\label{fig:chan}
\end{figure}

\begin{figure}
  \begin{center}
    \FigureFile(70mm,70mm){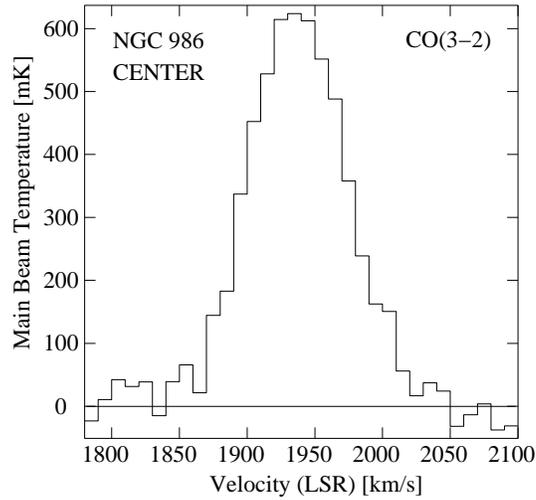}
  \end{center}
\caption{
ASTE CO(3--2) spectrum at the central position of NGC 986.
The effective spatial resolution (see Section 2) of the spectrum is $25''$
(HPBW). The peak temperature of approximately 0.6 K corresponds to 36 Jy beam$^{-1}$.
}\label{fig:ISPEC}
\end{figure}

The derived velocity channel maps of CO(3--2) emission in NGC 986
are displayed in figure \ref{fig:chan}. 
We find a strong concentration of CO emission toward the center with a large veocity
width (from 1865 to 2015 km s$^{-1}$; see also the CO(3--2) spectrum
at the central position in figure \ref{fig:ISPEC});
however, the extended emission along the SW to NE direction in some velocity channels
(from 1925 to 1985 km s$^{-1}$) is also evident.

\begin{figure}
  \begin{center}
  \FigureFile(150mm,150mm){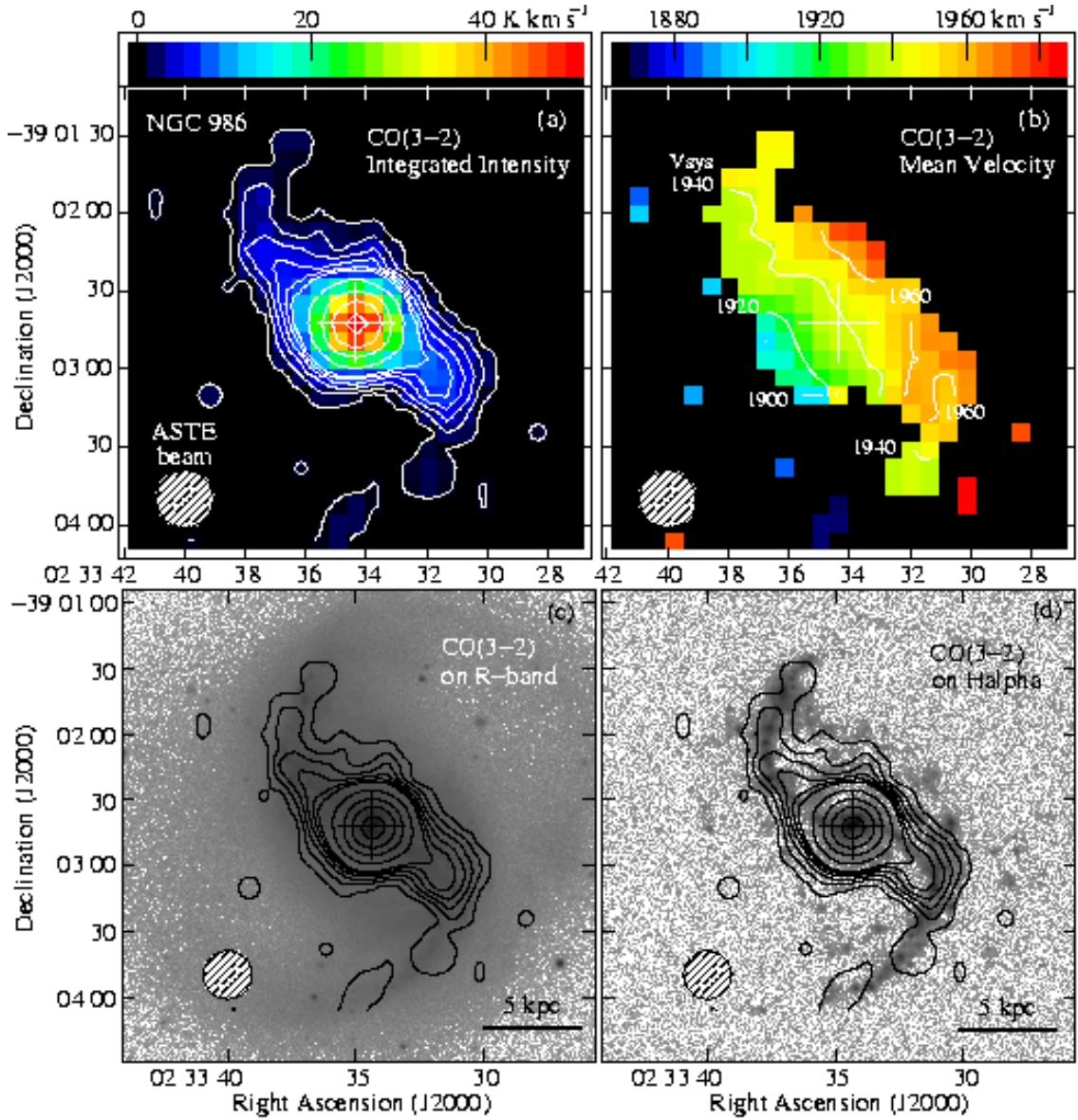}
  \end{center}
\caption{
(a) Velocity-integrated CO(3--2) intensity map of NGC 986, obtained
by the OTF mode using the SC345 receiver mounted on ASTE. 
The image size is 3$'$ $\times$ 3$'$ or 20 $\times$ 20 kpc
at a distance of 23 Mpc.
Contour levels are 1, 4, 7, 10, 13, 16, 30, 50, 70, \& 90 \%
of the peak intensity, 55.2 K km s$^{-1}$ or $3.35 \times 10^3$ Jy beam$^{-1}$ km s$^{-1}$.
The central cross indicates the position of the nucleus
(defined by the peak position of the NIR continuum from 2MASS/NED).
(b) Intensity-weighted mean velocity map of CO(3--2) in NGC 986.
Contours are labeled by the LSR velocity.
(c) Contour map of the CO(3--2) integrated intensity 
superposed on the R-band continuum image of NGC 986 (\cite{hameed1999}).
(d) Contour map of the CO(3--2) integrated intensity 
superposed on the continuum-subtracted H$\alpha$ image of NGC 986 (\cite{hameed1999})
}\label{fig:allmap}
\end{figure}

The 0th and 1st moment maps, i.e., a velocity integrated intensity map
and an intensity-weighted mean velocity map, respectively,  
are shown in figure \ref{fig:allmap}.
The peak position of CO(3--2) emission, fitted by the AIPS task IMFIT,
coincides well with the NIR peak within an accuracy of less than $1''$.

In addition to the strong CO condensation at the center,
we can clearly see extensions of the CO emission along the bar and outer spiral arms,
as observed in the R-band and H$\alpha$ images (\cite{hameed1999}).
The major axis of this gaseous bar seen in CO(3--2) emission is $\sim$ 2 arcmin or 14 kpc.
The nature of the gaseous bar will be discussed in section 4.

\subsection{Spatial coincidence between CO(3--2) and H$\alpha$ emissions}

In figure \ref{fig:allmap}(d), 
we find a good spatial coincidence between the overall distributions
of dense molecular gas traced by CO(3--2) 
and the massive star formation depicted by H$\alpha$
at a spatial scale of $\sim$ 3 kpc.
This is expected from the recent CO(3--2) observations that demonstrate
a tight correlation between CO(3--2) and H$\alpha$ luminosities (\cite{komugi2007})
thereby indicating an intimate association of the dense molecular medium with massive star formation
(see also \cite{yao2003}).
It is intriguing to study whether CO(3--2) emission also shows a {\it better} spatial
coincidence with the massive star forming regions as compared to the low-density gas
traced by the CO(1--0) emission. 
In fact, previous surveys of CO(3-2) and CO(1-0) lines
toward the centers of star-forming galaxies suggest that 
CO(3-2) luminosities show a {\it tighter} correlation 
with star-formation rates (traced by H$\alpha$ or FIR) than that of CO(1-0) luminosities
(\cite{komugi2007,yao2003}).
Future high-resolution low-J CO observations might be able to address this issue.

\subsection{Global CO(3--2) luminosity}

The total CO(3--2) flux of NGC 986 was measured using the AIPS task TVSTAT.
The mean CO(3--2) intensity was 4.5 K km s$^{-1}$ averaged
over 147 pixcels or $1.2 \times 10^8$ pc$^2$.
This gives a global CO(3--2) luminosity $L'_{\rm CO(3-2)}$
of $(5.4 \pm 1.1) \times 10^8$ in the unit of K km s$^{-1}$ pc$^2$. 
Note that the quoted error represents the systematic uncertainty, 
mainly due to the accuracy of main beam efficiency ($\pm$ 20 \%).
The derived CO(3--2) luminosity of NGC 986 is comparable with that of inner $5' \times 5'$
region of M 83 (\cite{muraoka2007}).

\subsection{CO(3--2)/CO(1--0) and CO(3--2)/CO(2--1) intensity ratios}
\label{ratio}

We compared the CO(3--2) intensity we obtained at the central position with the existing
CO(1--0) (\cite{aalto1991}; \cite{elfhag1996}) and CO(2--1) (\cite{aalto1995})
measurements. All the previous observations were made using the SEST 15 m telescope.
In order to compare our CO(3--2) data with the low-resolution 
($43''$ or $44''$ resolution) CO(1--0) intensities, we convolved our CO(3--2) cube
to the same spatial resolution ($44''$). 
The SEST CO(2--1) observations had a beam size
similar to that of the ASTE CO(3--2) observations, and no correction was made in the derivation
of a CO(3--2)/CO(2--1) line ratio.
The integrated intensities of various CO transitions
were then summarized (in Table 3).

The CO(3--2)/CO(1--0) integrated intensity ratio $R_{\rm 3-2/1-0}$ was 0.60 $\pm$ 0.13
at a spatial resolution of $44''$ or 5 kpc, and the CO(3--2)/CO(2--1) ratio $R_{\rm 3-2/2-1}$ was
0.67 $\pm$ 0.14 for a beam size of $\sim$25$''$ or $\sim$2.8 kpc.

The observed $R_{3-2/1-0}$ value, $\sim$0.6, is greater than those of the GMCs
in the disk regions of the Milky Way ($\sim$ 0.4, \cite{sanders1993}; $\sim$ 0.5, \cite{oka2007}) 
and M 31 ($\sim$ 0.3, \cite{tosaki2007}). 
Elliptical galaxies with very low-level star formation also show low ratios
($\sim$ 0.4, \cite{vila2003}).
Similar $R_{\rm 3-2/1-0}$ values are reported in the inner a few kpc
region of M 83 (0.6--0.7, \cite{muraoka2007}) and 
infrared luminous galaxies ($\sim$ 0.6, \cite{mauersberger1999}; \cite{yao2003}),
and dwarf starburst galaxies ($\sim$ 0.6, \cite{meier2001}). 
Some of nearby starburst galaxies or giant HII regions in galaxies exibit 
high $R_{3-2/1-0}$ values close to unity
(\cite{devereux1994, dumke2001, tosaki2007b}). 
The central molecular zone of the Milky Way (a size of $\sim$ 200 pc) 
also shows high $R_{\rm 3-2/1-0}$ ratio ($\sim$ 0.9, \cite{oka2007}).
Note that the apperture sizes/beam sizes of these line ratio measurements are 
a few 100 to a few kpc scales, and therefore mixture of multiple ISM components 
must be observed.

The observed line ratios in NGC 986 suggest a moderate excitation condition of the CO lines 
within the central a few kpc region of this galaxy.
For instance, $R_{\rm 3-2/2-1}$ of 0.7 suggest that the mean gas density $n_{\rm H_2}$ 
averaged over a $\sim$2.8 kpc region in diameter is
within a range of $10^3$ to $10^4$ cm$^{-3}$ based on an LVG model 
by assuming a kinetic temperature of 30 K (see Fig.\ 8 of \cite{hafok2003} for instance).
Additional mapping observations of low-J CO lines will be essential
to determine the spatial distribution of CO line ratios,
allowing us to impose constraints on the physical properties of ISM for various positions
in NGC~986.

\subsection{Molecular gas mass}

From our global CO(3--2) luminosity of NGC 986,
the total molecular gas mass was estimated to be 
\begin{equation}
M(\rm{H_2}) = 2.6 \times 10^9 
\left( \frac{\alpha_{\rm CO}}{2.9 \mbox{\ } M_\odot \mbox{(K km s$^{-1}$ pc$^2$)$^{-1}$ }  } \right) 
\left( \frac{R_{\rm 3-2/1-0}}{0.6} \right)^{-1} M_\odot.
\end{equation}
Here, we adopted a Galactic CO(1--0)-to-H$_2$ conversion
factor $\alpha_{\rm CO} = 2.9$ $M_\odot$ (K km s$^{-1}$ pc$^2$)$^{-1}$
(equivalent to $X_{\rm CO} = 1.8 \times 10^{20}$ cm$^{-2}$ (K km s$^{-1}$)$^{-1}$,
\cite{dame2001})) and 
a CO(3--2)/CO(1--0) ratio $R_{\rm 3-2/1-0}$ of 0.6 (see subsection \ref{ratio}).
This molecular gas mass is comparable to those in nearby gas-rich spiral galaxies
such as M 51 ($6 \times 10^9$ $M_\odot$, \cite{kuno1995paperII}) 
and M 83 (3--4 $\times 10^9$ $M_\odot$,
\cite{crosthwaite2002}; \cite{lundgren2004}).

\subsection{Star formation efficiency}

The star formation efficiency (SFE), a star formation rate per unit gas mass, 
is then $L_{\rm FIR}/M({\rm H_2}) = 14$ $L_\odot/M_\odot$.
Note that both $L_{\rm FIR}$ and $M({\rm H_2})$ were measured over the almost entire region
of this galaxy (i.e., determined for the same area).
This SFE in NGC 986 is indeed close to the SFEs of nearby isolated or field spiral galaxies,
$L_{\rm FIR}/M({\rm H_2}) \sim 3-5$ $L_\odot/M_\odot$ (\cite{young1996}).
Note that \citet{young1996} assumed 
an $X_{\rm CO} = 2.8 \times 10^{20}$ cm$^{-2}$ (K km s$^{-1}$)$^{-1}$;
if we recalibrate the SFEs using the same $X_{\rm CO}$ value as for NGC 986 (\cite{dame2001}),
the SFEs for nearby isolated or field spirals will be $\sim 5-8$ $L_\odot/M_\odot$, 
which are more closer to that of NGC 986.
On the other hand, the observed SFE in NGC 986
is much lower than those in local IR luminous galaxies, 
$L_{\rm FIR}/M({\rm H_2}) \sim 120$ $L_\odot/M_\odot$ 
(a mean value computed from Table 2 of \cite{yao2003}).
Here \citet{yao2003} determined the $X_{\rm CO}$ of their sample 
to be $2.7 \times 10^{19}$ cm$^{-2}$ (K km s$^{-1}$)$^{-1}$, 
about 1/10 of a typical Galactic CO-to-H$_2$ conversion factor.

Another direct quantity related to SFE is the FIR to CO(3--2) luminosity ratio 
$L_{\rm FIR}/L'_{\rm CO(3-2)}$,
a measure of star formation rate per unit dense-gas mass.
We find the $L_{\rm FIR}/L'_{\rm CO(3-2)}$ ratio of 67 $L_\odot$/(K km s$^{-1}$ pc$^2$)
in NGC 986. This is comparable to the $L_{\rm FIR}/L'_{\rm CO(3-2)}$ values
found in local IR luminous galaxies, a few 10 $L_\odot$/(K km s$^{-1}$ pc$^2$),
that was calculated from \cite{yao2003},
yet still much smaller than Ultra/Hyper luminous IR galaxies in the early universe,
such as the submillimeter galaxy MIPS J142824.0+352619 at $z$ = 1.3, 
showing the $L_{\rm FIR}/L'_{\rm CO(3-2)}$ of 
264 $\pm$ 84 $L_\odot$/(K km s$^{-1}$ pc$^2$) (\cite{iono2006}).

\subsection{Kinematics of dense molecular gas}

We find a clear velocity gradient in the mean-velocity map (figure \ref{fig:allmap}(b)).
We therefore determined the kinematical parameters of the gas disk, i.e., 
the dynamical center, systemic velocity, 
the position angle of the major axis, and inclination angle of the disk 
by a least-squares fitting of the intensity-weighted isovelocity field 
to a circular rotation model.
The AIPS task GAL was used for this analysis. 
The fitting was made within a radius of 1$'$,
where the observed velocity field seems to be dominated by a circular motion.
In general, strong non-circular motions are expected in the central
regions of barred galaxies, but it is not clear in our map; it could be 
probably due to the insufficient spatial resolution of our observations.

The dynamical center coincides well with the nucleus position
determined by the NIR (2MASS) peak within an error of a few arcsec.
The systemic velocity (LSR) was determined to be 1942 $\pm$ 10 km s$^{-1}$
(or corresponding to a heliocentric velocity of 1957 $\pm$ 10 km s$^{-1}$).
This agrees well with the CO(1--0) result (\cite{elfhag1996})
and is close to the HI velocity measurement (\cite{roth1994}).
See Table 2 for comparison.
The position angle of the major axis and the inclination angle of the disk
were estimated to be 127$^\circ$ (from north to east) and 
37$^\circ$ (0$^\circ$ is face-on), respectively.
These kinematically determined angles are consistent 
with the previously reported isophotal values based on 
optical and NIR broad band images of NGC 986
(\cite{tully1988}; \cite{RC3}; \cite{Jarrett2003}).
The comparison of these angles is shown in Table 2.

\begin{figure}
  \begin{center}
    \FigureFile(70mm,70mm){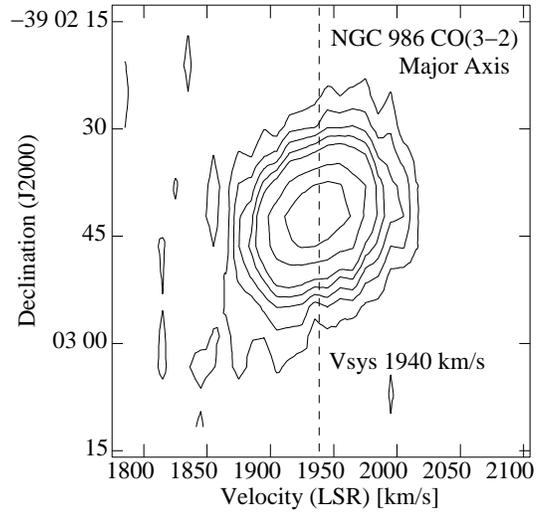}
  \end{center}
\caption{
ASTE CO(3--2) position-to-velocity (PV) diagram along the major axis 
(P.A.\ = 127$^\circ$; Table 2) of NGC 986.
Contour levels are 2, 4, 6, 8, 10, 15, 20, and 25 $\sigma$, where
1 $\sigma$ = 25 mK on the $T_{\rm mb}$ scale or 1.5 Jy beam$^{-1}$. 
}\label{fig:PV}
\end{figure}

A position-to-velocity map (PV map) was generated along the determined major axis
(P.A.\ = 53$^\circ$), as shown in figure \ref{fig:PV}. 
The velocity gradient along the major axis was found to be 
$\sim$10 km s$^{-1}$ arcsec$^{-1}$ from the figure;
however, but we require much higher angular resolution measurements of the PV map,
using SMA for instance, in order to determine the inner rotation curve of NGC 986.

\section{Discussion: a large ($\sim$14 kpc) gaseous bar filled with dense molecular medium}

Our CO(3--2) image revealed the presence of 
a large ($\sim$ 14 kpc) gaseous bar
filled with dense molecular medium
along the dark lanes observed in the optical images.
This is the largest ``dense-gas rich bar'' known to date;
previous large scale CO(3--2) imaging observations of nearby galaxies
revealed that a centrally concentrated CO(3--2) morphology of around a few kpc scale 
is very common (\cite{hurt1993}; \cite{mauersberger1996}; \cite{ib2001}; \cite{dumke2001}; 
\cite{ib2003}).
Some gas-rich barred spiral galaxies such as 
NGC 6946 (\cite{ib2001}; \cite{walsh2002}), 
and M 83 (\cite{ib2001}; \cite{bayet2006}; \cite{muraoka2007}), 
exhibit a wide spread CO(3--2) emission over the disk region; 
however, no such instances of large ($\sim$10 kpc scale) 
dense molecular gas bar have been reported thus far.
Therefore, it is indeed a surprise to discover a 14 kpc long gaseous
bar observed in CO(3--2) emission among nearby spiral galaxies.
Note that this is even true in the case of bars observed in low-J CO lines. 
Some galaxies such as NGC 1530 (\cite{downes1996}), 
UGC 2855 (\cite{hutte1999}), and NGC 7479 (\cite{sempere1995}) 
do exhibit a continuous bar, observed in CO(1--0) and/or CO(2--1) emission, 
with a length of $\sim$10 kpc; however, this is a rare phenomenon.

The dense molecular medium in the bar region could be transported
to the central region of NGC 986 within a short time scale ($\sim$ a dynamical time scale)
due to the strong shock along the bar 
(e.g., see \cite{wadahabe1992} and references therein). 
This could be a major reason why
we hardly observe such a gas-rich bar in galaxies (e.g., \cite{hutte1999}). 
Therefore, the discovered dense gas bar in NGC 986 must be a huge reservoir
of possible ``fuel'' for future starbursts in the central region, 
and we suggest that the star formation process 
in the central region of NGC 986 could still be in a growing phase.

\vspace{0.5cm}

We are grateful to the referee, Susanne H\"uttemeister, 
for her careful reading of the manuscript and useful comments.
We would like to acknowledge all the members involved with the ASTE team 
for their great efforts in the ASTE project. This study was financially supported
by MEXT Grant-in-Aid for Scientific Research on Priority Areas No.\ 15071202.
Observations with ASTE were carried out 
remotely from Japan by using NTT's GEMnet2 and its partner 
R\&E (Research and Education) networks,
which are based on AccessNova 
collaboration of University of Chile,
NTT Laboratories, and National Astronomical 
Observatory of Japan.



\clearpage

\clearpage


\begin{table*}
\begin{center}
Table~1.\hspace{4pt}Sample galaxies of ADIoS project.\\
\begin{tabular}{ccccc}
\hline \hline
Object    & Map size        & Mode$^{\dagger}$ & Reference           & Complementary CO(1--0) data \\ 
\hline
M 83    & $5' \times 5'$  & PS     & \cite{muraoka2007} & NRO 45 m (\cite{kuno2007}) \\
a GMA in M 31 & $1'.5 \times 1'.5$ & PS  & \cite{tosaki2007}  & NRO 45 m (\cite{tosaki2007}) \\
NGC 986 & $3' \times 3'$  & OTF    & this work & --- \\
NGC 253 & $9' \times 3'$  & OTF    & Nakanishi et al.\ (in prep.) & NRO 45 m (\cite{sorai2000n253}) \\
NGC 604 (in M33) & $5' \times 5'$  & OTF & \cite{tosaki2007b}  & NRO 45 m (\cite{tosaki2007b}) \\
\hline
\end{tabular} \\
\label{table-adios}
\end{center}
{\footnotesize
$^{\dagger}\hspace{4pt}$
PS = position switching, OTF = on-the-fly mapping.
}
\end{table*}


\begin{table*}
\begin{center}
Table~2.\hspace{4pt}Properties of NGC 986.\\
\label{table:param}
\begin{tabular}{llll}
\hline \hline
\multicolumn{2}{l}{Parameter} & Value & Reference \\
\hline
\multicolumn{2}{l}{Distance} & 23.2 Mpc & (1) \\
\multicolumn{2}{l}{Scale} & 112 pc/$''$ &     \\
\multicolumn{2}{l}{Morphology} & (R$'_1$)SB(rs)b & (2) \\
Position of nucleus & R.A.(J2000) & $02^h 33^m 34^s.35$ & (3) \\
                  & Decl.(J2000) & $-39^{\circ} 02' 42''.2$ &  \\
\multicolumn{2}{l}{Systemic velocity (LSR)$^{*}$} & 1969 km s$^{-1}$ & (4) \\
\multicolumn{2}{l}{                                   } & 1945 km s$^{-1}$ & (5) \\
\multicolumn{2}{l}{                                   } & 1942 $\pm$ 10 km s$^{-1}$ & this work \\
\multicolumn{2}{l}{Inclination (0$^\circ$ is face-on)} & 42$^{\circ}$ & (1) \\
\multicolumn{2}{l}{}            & 41$^{\circ}$ & (6) \\
\multicolumn{2}{l}{}            & 55$^{\circ}$ & (7) \\
\multicolumn{2}{l}{}            & 37$^{\circ}$ & this work \\
\multicolumn{2}{l}{Position angle (from north to east)} & 150$^{\circ}$ & (6) \\
\multicolumn{2}{l}{}               & 127$^{\circ}$ & (7) \\
\multicolumn{2}{l}{}               & 127$^{\circ}$ & this work \\
\multicolumn{2}{l}{Global FIR luminosity $L_{\rm FIR}$} & $3.5 \times 10^{10} L_{\odot}$ & (8) \\
\multicolumn{2}{l}{Global CO(3--2) luminosity $L'_{\rm CO(3-2)}$} & $(5.4 \pm 1.1) \times 10^8$ K km s$^{-1}$ pc$^2$ $^{\dagger}$& this work \\
\multicolumn{2}{l}{Molecular gas mass $M({\rm H_2})$} & $2.6 \times 10^{9} M_{\odot}$ $^{\ddagger}$ & this work \\
\multicolumn{2}{l}{Star formation efficiency $L_{\rm FIR}/M({\rm H_2})$ } & 14 $L_\odot/M_\odot$ & this work \\
\multicolumn{2}{l}{Star formation efficiency $L_{\rm FIR}/L'_{\rm CO(3-2)}$ } & 67 $L_\odot$/(K km s$^{-1}$ pc$^2$) & this work \\
\hline
\end{tabular}\\
\end{center}
{\footnotesize
$^{*}\hspace{4pt}$
To obtain the heliocentric velocity, add 15 km s$^{-1}$ to the velocity in LSR. \\
$^{\dagger}\hspace{4pt}$
The quoted error represents systematic uncertainty, mainly due to the accuracy 
of the main beam efficiency (20 \%). \\
$^{\ddagger}\hspace{4pt}$
Adopting a Galactic CO-to-H$_2$ conversion
factor $\alpha_{\rm CO} = 2.9$ $M_\odot$ (K km s$^{-1}$ pc$^2$)$^{-1}$
or $X_{\rm CO} = 1.8 \times 10^{20}$ cm$^{-2}$ (K km s$^{-1}$)$^{-1}$ (\cite{dame2001}) 
and a CO(3--2)/CO(1--0) ratio $R_{\rm 3-2/1-0}$ of 0.6 (see section \ref{ratio}). \\

References: 
(1) \cite{tully1988}; 
(2) \cite{buta1995};
(3) 2MASS EXTENDED OBJECTS, Final Release (2003)/NED; 
(4) HI, \cite{roth1994};
(5) CO(1--0), \cite{elfhag1996};
(6) RC3 (de Vaucouleurs et al.\ 1991);
(7) \cite{Jarrett2003}/NED; 
(8) \cite{sanders2003}, rescaled to the value for the distance we adopted here.
}
\end{table*}


\begin{table*}
\begin{center}
Table~3.\hspace{4pt}Multi transition CO intensities at the center of NGC 986.\\
\begin{tabular}{llllll}
\hline \hline
Transition & Integrated intensities                & Telescope & Beam efficiency & Beam size & Reference \\
           & $\int T_{\rm mb} dv$ [K km s$^{-1}$] &           & $\eta_{\rm mb}$ & [$''$]    &     \\
\hline
CO(1--0)    & 33.4 $\pm$ 0.6 & SEST 15 m & 0.7  & 43 & \cite{aalto1991} \\
CO(1--0)    & 36.2 $\pm$ 0.7 & SEST 15 m & 0.7  & 44 & \cite{elfhag1996} \\
CO(2--1)    & 82.4 $\pm$ 1.2 & SEST 15 m & 0.5  & 24 & \cite{aalto1995} \\
CO(3--2)    & 55.2 $\pm$ 1.1 $\pm$ 11$^{\dagger}$ & ASTE 10 m & 0.6  & 25 & this work \\
CO(3--2)    & 21.3 $\pm$ 0.33 $\pm$ 4.3$^{\dagger}$ & ASTE 10 m & 0.6 & 44 & this work (convolved) \\
\hline \hline
\multicolumn{3}{l}{} & Ratio & Beam size &  \\
\multicolumn{3}{l}{} &       & [$''$] & [kpc]  \\
\hline
\multicolumn{3}{l}{CO(3--2)/CO(1--0) integrated intensity ratio $R_{3-2/1-0}$} & 0.60 $\pm$ 0.13 & 44 & 5.0 \\
\multicolumn{3}{l}{CO(3--2)/CO(2--1) integrated intensity ratio $R_{3-2/2-1}$} & 0.67 $\pm$ 0.14 & 25 & 2.8 \\
\hline
\end{tabular} \\
\end{center}
{\footnotesize
$^{\dagger}\hspace{4pt}$the first error corresponds to the S/N of the spectrum 
(i.e., random error only),
and the second error is obtained from the systematic error, 
mainly due to the accuracy of the main beam efficiency (20 \%).
}
\end{table*}

\end{document}